\documentclass[runningheads]{llncs}
\usepackage{graphicx}
\begin{document}
\title{Operational Research Literature as a Use Case for the Open Research Knowledge Graph}
\titlerunning{OR Literature as a Use Case for the ORKG}
%
\author{Mila Runnwerth\inst{1} \and
Markus Stocker\inst{1}\orcidID{0000-0001-5492-3212} \and Sören Auer\inst{1}\orcidID{0000-0002-0698-2864}}
\authorrunning{Runnwerth, Stocker, Auer}
%
\institute{{TIB - Leibniz Information Centre for Science and Technology, Welfengarten 1B, 30167 Hannover, Germany } \\
\url{http://www.tib.eu} \\
\email{\{firstname.surname\}@tib.eu}}
\maketitle              
\begin{abstract}
The Open Research Knowledge Graph (ORKG) provides machine-actionable access to scholarly literature that habitually is written in prose. Following the FAIR principles, the ORKG makes traditional, human-coded knowledge findable, accessible, interoperable, and reusable in a structured manner in accordance with the Linked Open Data paradigm. At the moment, in ORKG papers are described manually, but in the long run the semantic depth of the literature at scale needs automation. Operational Research is a suitable test case for this vision because the mathematical field and, hence, its publication habits are highly structured: A mundane problem is formulated as a mathematical model, solved or approximated numerically, and evaluated systematically. We study the existing literature with respect to the Assembly Line Balancing Problem and derive a semantic description in accordance with the ORKG. Eventually, selected papers are ingested to test the semantic description and refine it further.

\end{abstract}
\section{Introduction}
Today's scholarly communication behaviour and logistics is still defined by centuries of printed document culture. Although there is progress by transforming journals into digital article repositories that, in principle, provide access to the content at all times and irrespective of a researcher's location, the nature of an article itself has not changed: The investigated hypothesis, the used methodology, the experiment, and the outcome are written in prosaic form; the final document is usually published for no other purposes than reading, seemingly optimised for human cognition.

The Open Research Knowledge Graph (ORKG)~\cite{Jaradeh2019} questions the “paradigm of document-centric scholarly information communication”~\cite{Auer2018}. It aims at transforming research literature into structured, machine-actionable data in order to represent and express information through semantically rich, interlinked knowledge graphs. Similarly to DBpedia, a prosaic knowledge source is transformed according to Linked Open Data standards~\cite{Auer2007}. Users are enabled to compare papers, discover patterns across methods or disciplines, or get a structured overview in a chosen context.

The main use cases of the ORKG's beta version\footnote{\url{https://orkg.org/}} are article search, a machine-actionable, semantic representation, and especially paper comparison as introduced in~\cite{Oelen2019}. To date, it indexes about 400 research articles. More than half are assigned to the subject cluster \emph{Physical Sciences \& Mathematics}. 

\subsection{Operational Research as a Use Case from Mathematics}
The structural science mathematics provides particularly suitable content for the ORKG: Its published prose is clear and dense from a linguistic point of view. However, as~\cite{Brack2020} have shown the high degree of abstraction in mathematics makes a conceptualisation consisting of the categories \emph{process}, \emph{method}, \emph{material}, and \emph{data}, which have been adapted to empirical sciences, inexpedient. The ORKG is not limited to this model but its feature, the abstract annotator, has been shown to be out of its depth with regard to mathematics.
The applied mathematical science of operational research (OR) combines the rather abstract fields of combinatorics and numerical analysis with mundane research questions from economics. In favour of this study, we narrowed the topic down to the optimisation problem of \emph{Assembly Line Balancing}. Its name derives from mass production where the intricate logistics for paced manufacturing assembly lines have to be organised efficiently, i.~e. optimally. The Assembly Line Balancing Problem (ALBP) and its variations are not only well-covered in scholarly literature but also provide an abundance of structured overviews of exact and heuristic algorithms or benchmarks thereof. Thus, the research literature about ALBPs is an appropriate use case for the ORKG.
In a first step, we choose literature reviews and articles that suggest minor optimisations to existing methods, which are compared to each other. Then, we suggest a semantic description that covers the content of the collection. It will serve as a prospective template for the ORKG. Furthermore, we will look for elements and patterns in the papers that are suited for automatic extraction in the future. Third, we ingest those literature reviews or articles that are published under an eligible licence, i.~e. a CC-BY, CC-BY-SA, or arXiv's \emph{Non-exclusive licence to distribute} into the ORKG. During this intellectual step, we will test and refine the proposed data model.  Finally, we consider the representations and comparisons of the scholarly contributions in the ORKG and discuss its added value for researchers.

\section{A Template for the Assembly Line Balancing Problem}


Scholarly research of the ALBP can be traced back to the 1960s when it was shown to be a NP-hard combinatorial optimisation problem~\cite{Gutjahr1964}. Since then scientists work on the sophistication of the mathematical model, exact algorithms for defined special cases or heuristic algorithms in order to find optimal solutions in adequate time. Recently, several reviews have been published to benchmark the stated mathematical model, exemplary data scenarios~\cite{Chaves2007} or performances of the selected methods. We chose to set a focus on these reviews at first, but most publications were not openly accessible or free to be reused in the ORKG according to their respective licences. Eventually, articles introducing a new model statement for the ALBP or a heuristic to solve it were also considered. The collection comprised 28 topically relevant papers of which eight provide openly accessible preprints on \emph{arXiv}\footnote{\url{arxiv.org}}. These were manually ingested into the ORKG with varying degrees of thoroughness (cf. Section~\ref{EnteringALBPLiteratureIntoTheORKG}): From the single statement of the research problem to detailed descriptions of the algorithms and data sets that were applied\footnote{An exemplary comparison of three selected papers can be found at \url{https://www.orkg.org/orkg/comparison?contributions=R12018,R12059,R12193}.}.

The collection of research articles was organised in the open source reference management system Zotero\footnote{\url{https://www.zotero.org/}}, also including documented experiences of the whole process. 

\subsection{A Semantic Model to Reflect ALBP Research}
\label{SemanticModelToReflectALBPResearch}

The ORKG's performance depends on a data model that is well-tuned to the content it is supposed to represent. That means expert knowledge in both the considered field and data modelling is required. Authors who possess the domain knowledge may not be able to squeeze it into the RDF scheme of the ORKG because there is no or little expertise in knowledge engineering. Data curators on the other hand may struggle with the proper in-depth indexing of the latest research knowledge. The ORKG's flexibility is an advantage because it allows almost limitless adaptions to describing papers by reusing existing concepts (mostly entries by former contributors) and relations but also by introducing new ones. The default schema stems from the comprehension of empirical sciences: A method is applied to a defined research question. This application causes a process that involves material to be observed or changed. Meanwhile observational data is collected and eventually evaluated in order to prove or disprove a hypothesis constructed prior to the experiment. 

In operational research in general and with respect to the ALBP in particular, there is also a rather standardised development that can be represented by a data model: The practical problem is formulated as a mathematical model or programme. Depending on the choice of the model, there is a toolbox of direct or heuristic algorithms to yield an exact (or approximate) solution to the model. Usually, in scholarly literature either a new variant of the ALBP is stated and the derived model is traced back to established methods or a new or rather slightly modified method is tested against known methods to solve the same problem. Thus, we conclude that most research papers about the ALBP are comprised of the elements listed in Table~\ref{ALBPRelations}. 

\begin{table}[]
    \centering
    \caption{A semantic model translating the OR research process into the ORKG scheme. The arrows connote a hierarchical descent, the asterisks connote a newly introduced property.}
    \begin{tabular}{l l l}
        \textbf{OR Term} & \textbf{ORKG Properties} & \textbf{Example} \\
        Name of the optimisation problem & Has research & ALBP \\
        Model / programme & Has approach $\rightarrow$ & MIP \\
        & Has model$^*$ & \\
        Exact method & Has method $\rightarrow$ & Branch \& bound \\
        & Has exact solution method$^*$ & \\
        Heuristic & Has method $\rightarrow$ & Tabu search \\
        & Has heuristic$^*$ & \\
        Instance data set & Has instance$^*$ & Roszieg \\
        Programming language & Has implementation $\rightarrow$ &  C, GCC 3.4.0 \\
        & Has programming language$^*$ &  \\
        System specifications & Has implementation $\rightarrow$ & Athlon 64 X2 4400 \\
        & Has system specification$^*$ & \\
        Performance & Has performance$^*$ & $\mathcal{O}(n\log{}n)$, 0,2~ms
    \end{tabular}
    \label{ALBPRelations}
\end{table}

\emph{Has Performance} contains the results that depend on the method that is applied, the graph the algorithm is applied on, and the specification of the implementation and system. Thus, it is semantically interlinked with other elements. 

If the suggested structure in Table~\ref{ALBPRelations} proves valid, it can be cast into a topic-specific template on its own in order to facilitate highly consistent knowledge graphs of further relevant papers independent of the curator.

\subsection{Entering ALBP Literature into the ORKG}
\label{EnteringALBPLiteratureIntoTheORKG}

After careful study and annotation of the eight papers from arXiv, we entered the data into the ORKG. In the first of three steps of the procedure, the formal metadata can be automatically ingested via DOI\footnote{\url{https://www.doi.org/}} or a BibTeX entry. There is an additional fallback option to enter the formal metadata manually. Since the preprint repository arXiv does not provide a DOI for its documents, we chose BibTeX entries for the import.

In the second step, the document is classified by subject. The ORKG's specified, hierarchical classification does currently not allow for several attributions. Hence, when assigning a single subject, a multidisciplinary field such as operational research is prone to inconsistencies with respect to its main focus in the respective paper or the curator. We chose to consistently assign the collection to Applied Mathematics $\rightarrow$ Numerical Analysis \& Computation, although several other closely related fields would have been adequate as well, for example Engineering $\rightarrow$ Operations Research (and more). However, OR being predominantly a multidisciplinary subject involving mathematics, computer science, and economics, engineering seemed too misleading for a semantically sound assignment.


The curator may choose between several templates; the template called \emph{Research Problem} is closest to the data model as suggested in Section~\ref{SemanticModelToReflectALBPResearch}. The template provides the field \emph{Has research} where keywords can be chosen from the suggested list or entered manually. Each entry is added to a bag-of-words and, thus, will be provided for autocompletion further on. This unrestricted freedom leads to a number of challenges. We struggled with typos (e.~g. 'optimsation') as it was not immediately obvious how to correct these. Moreover, the same word was included in its American and British form, respectively, i.~e. 'optimization' and 'optimisation'. We plan to add some functionality to ORKG to semi-automatically interlink such surface forms as they are describing the same concept. An underlying controlled vocabulary with an additional feature to enter free text would avoid wreaking havoc in the bag-of-words. A user entering 'optimisation problem' may thus be faced with four versions of which one contains a typo and two are identical.

Further predefined fields are \emph{Has evaluation}, \emph{Has approach}, \emph{Has method}, \emph{Has implementation}, \emph{Has result}, \emph{Has value}, and \emph{Has metric}. Not all of these semantic relations make sense for describing a mathematical paper, or rather, they lack distinctive accuracy, e.~g. when does an approach become a method; or do we mean the outcome of the algorithm or its performance when stating the result? Yet, the relevant semantic units of an OR paper can be transferred and amended easily. 

Each field contains further fields in turn that may be annotated and indexed. And as a last resort, new relations can be introduced on every hierarchical level. The OR terms introduced in Table~\ref{ALBPRelations} were mapped by employing existing relations and introducing new ones (marked by an asterisk in the table). After leaving the hierarchical top level which is edited in the main browser window, every edit thereafter is conducted in a small overlay window. So while modelling, there is no visual aid where the description process is hierarchically taking place at the moment. However, we made it a habit to describe the top level first, save the description, such that the visualisation of the graph is available. From there, refinement is more accessible.

The eight papers are not consistently described in this fashion because each paper gave reason to a refinement iteration of previous graphs. Thus, after each paper, there are (or should be) well-documented, retroactive modifications to each graph representing a paper. Again, this inevitably leads to inconsistencies even among papers that are ingested by the same curator. Another critical observation is our choice of terms: General denominations such as \emph{Has model}, \emph{Has instance}, or \emph{Has performance} could mean completely different things in another context. Even between OR researchers these terms might not be semantically tight enough to guarantee frictionless communication. Hence, the relations are prone to cross-contextual use that might make the otherwise carefully created model fuzzy.

\begin{figure}
\centering
\includegraphics[width=\textwidth]{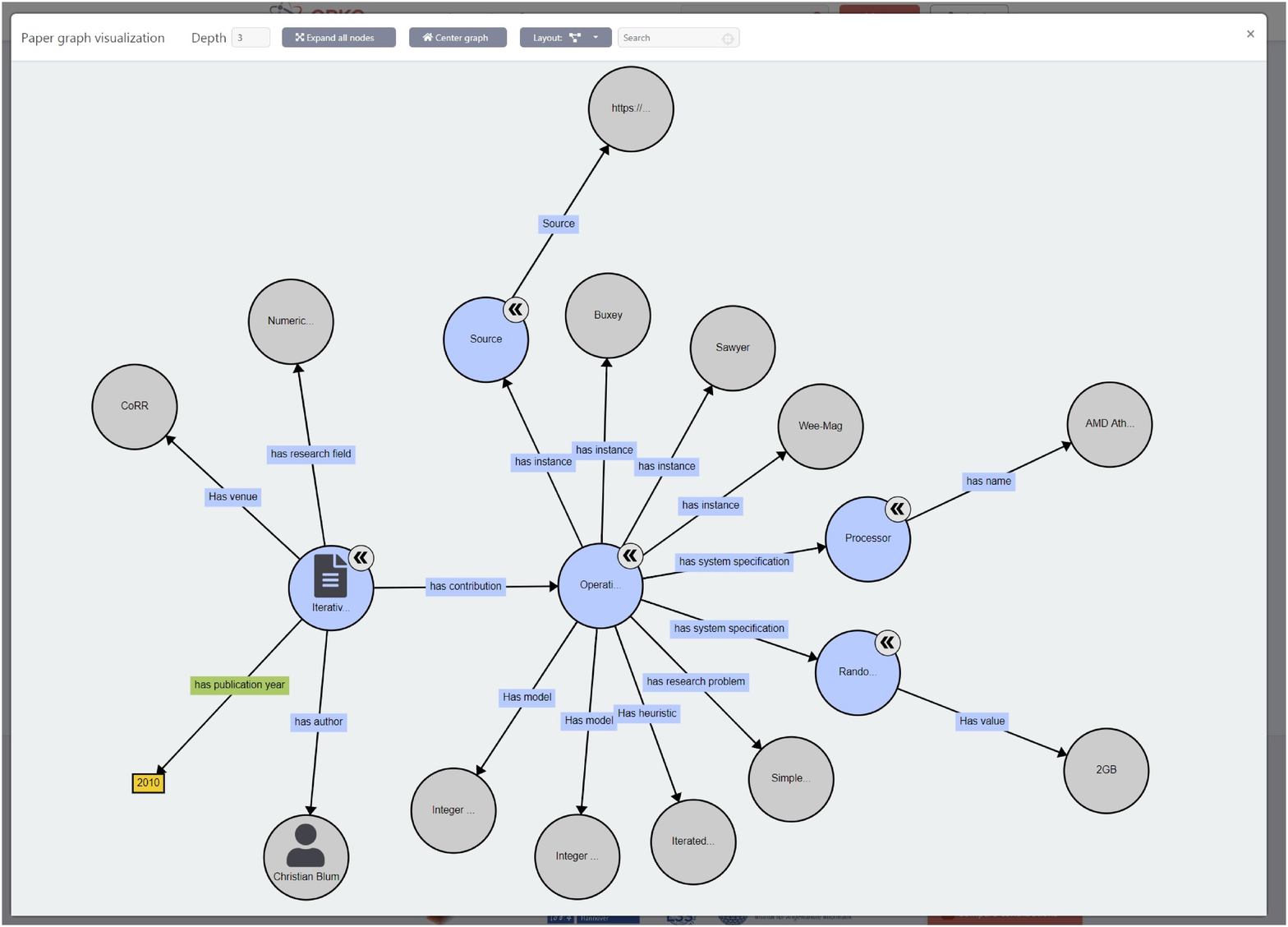}
\caption{Visualisation of a paper's knowledge graph.} 
\label{orkgvis}
\end{figure}

In theory, a paper can be thoroughly represented by modelling each sentence as Linked Data, at arbitrary granularity. Again, in agreement with another field expert from biochemistry, we concluded: when to finish the indexing procedure is at the margin of discretion. Of course, the ORKG's crowd-sourcing philosophy allows and even demands for further refinement by others or at a later stage. Thus, a knowledge graph is never truly complete, especially if dynamic data such as citations will be taken into account in the future. An exemplary paper description is shown in Figure~\ref{orkgvis}.

\section{Conclusion and Further Work}

OR is a suitable test case for the ORKG, because the topic itself and the appropriate publication habits are highly structured and can easily be mapped to the default data schema already provided by the ORKG. However, on the basis of this study we tackle several general and subject-specific further improvements in the future:

\begin{itemize}
    \item Creating checklists and guidelines to define both minimum requirements and a gold standard for a paper's knowledge graph.
    \item Underlying a general, and for templates a subject-specific, vocabulary with moderated editing workflows. Also, the resources should be displayed alphabetically or by assigned relevance instead of a last-in-first-out fashion in the tabular view.
    \item Linear user guidance for generating a skeleton data set and visual support for the refinement.
    \item Similarly described papers with identical or semantically close descriptions should yield similarity.
    \item The selected papers met our expectations of being highly structured and easy to parse for the defined information patterns. They are well suited for a pilot study of automated extraction for information framing a basic knowledge graph. Automated indexing where scientific literature is indexed with terms of well-maintained thesauri like the German Authority File or automatically classified with the Mathematics Subject Classification (MSC)\footnote{\url{http://msc2010.org/mediawiki/index.php?title=MSC2010}} may provide a first draft to be ingested into the ORKG\cite{Kasprzik2020}. 
    \item The aforementioned MSC would provide the obvious classification backbone for contributions from the mathematical sphere. The extremely confined example of the ALBP suggests that this would not only call for 63 template schemas for each top level class but at least 5.000 refinements accounting for MSC's subclasses. However, with this first experiment we cannot estimate the structural synergies between classes. We rather expect, given a wisely chosen sample that future work might result in a manageable number of mathematical templates with few extensions for the subclass topics. An example are the MSC classes 44 and 45 covering ordinary and partial differential equations, respectively. Even if they turn out to differ minutely in their ORKG template, these differences will be provided for in other templates, e. g. (numerical) analysis.
    \item Since the ORKG follows a crowdsourcing philosophy, seeking support from and collaborate with further projects in the field of mathematical knowledge engineering guarantees high quality and integrity of the data and its community-curated modelling. Critical exchange with the researchers of \textit{MathDataHub} is established\cite{Bercic2019}, but projects like swMATH, a database for mathematical software, should be considered more closely\cite{Chrapary2017}.
\end{itemize}
%
%
%
%

\end{document}